\documentclass[conference]{IEEEtran}

\usepackage[cmex10]{amsmath}

\usepackage{epsfig,amssymb,amsfonts,url}

\begin{document}

\title{Network protocol scalability via a topological Kadanoff transformation}

\author{Costas~Constantinou,~\IEEEmembership{Member,~IEEE,}
        and~Alexander~Stepanenko 

\thanks{Both authors are with the Department of Electronic, Electrical and Computer Engineering, University of Birmingham, Edgbaston, Birmingham, B15 2TT, UK e-mail: \{c.constantinou, a.stepanenko\}@bham.ac.uk}
}

\markboth{PhysComNet 2008}%
{Constantinou \MakeLowercase{\textit{et al.}}: Network protocol scalability}

\IEEEspecialpapernotice{(Invited Paper)}

\maketitle

\begin{abstract}

A natural hierarchical framework for network topology abstraction is presented based on an analogy with the Kadanoff transformation and renormalisation group in theoretical physics. Some properties of the renormalisation group bear similarities to the scalability properties of network routing protocols (interactions). Central to our abstraction are two intimately connected and complementary path diversity units: simple cycles, and cycle adjacencies. A recursive network abstraction procedure is presented, together with an associated generic recursive routing protocol family that offers many desirable features.
\end{abstract}


\section{Introduction}

\IEEEPARstart{T}{he} notion of routing protocol scalability is familiar to all researchers in the field of communication networks. Scalability is a desirable property of a network, but is hard to define rigorously. A router, or even a network whose performance improves after adding hardware (memory, switching capacity, bandwidth, etc.), monotonically with the added capacity, is usually referred to as a scalable system. An algorithm, or networking protocol, is said to scale if it is suitably efficient and practical when applied to a large number of participating nodes in the case of a distributed system.

In the context of distributed network routing protocols, scalability addresses the manner in which the following grow as the size of the network increases:

\begin{itemize}
\item the communication \& control overhead bandwidth consumption in collecting, processing and disseminating information on which to base forwarding decisions,
\item the computational complexity (and associated processing time at each node) in making forwarding decisions, and
\item the address space and associated size of routing tables
\end{itemize}

How well we can perform the above tasks depends on both how efficiently we can encode {\em all} the network paths, as well how efficiently we can select a specific subset from this multiplicity.

In theoretical physics, renormalisation group refers to a mathematical technique which uses a {\em natural hierarchy of scales} of interactions between increasingly larger coarse-grained ``blocks" governing the system properties at each scale. Blocks at every level are constructed from blocks at the next finer-grained level below. The class of interactions between such blocks amenable to a renormalisation group treatment is such that the effective interaction type between blocks is preserved at all levels of coarse-graining. For example, in studying magnetism the blocks in question are magnetic domains and the interaction is the dipole interaction between them. One can consider blocks of such domains and then blocks of blocks, and so on. The interaction between such coarse-grained blocks is of the same type (i.e. a dipole one) as at the microscopic level.

The above hierarchy of scales is the key idea which enables us to achieve a ``scalable" (or efficient) description framework for a system. It has been long recognised \cite{hierarchicaltopology} that hierarchical topology abstraction and summarisation are significant in ensuring the scalability of routing protocols in communication networks, but only concerning the compactness of the addressing scheme employed.

This paper, introduces a {\em natural hierarchy of topological scales} and then uses the renormalisation group as an inspiration to construct topological objects and interactions between them in such a manner as to preserve the type of interaction at all levels in the topological hierarchy.

At the most basic level, the atomic topological objects are vertices (nodes) in a graph-theoretic representation of a network and atomic interactions are edges (links) between vertices. The renormalisation group allows us to describe the interaction between distant parts of the system through knowledge of the corresponding interaction at the micro-level. In our case, we are attempting to describe connectivity between distant parts of a network through knowledge of progressively local connectivity down all the levels of the topological hierarchy, including the microscopic level. This hierarchy can be used for many purposes, including augmenting it with routing protocol rules, thus introducing a new class of routing protocols that have desirable scalability properties built-in by construction.

\section{Fundamental Topological Units of Path Diversity}

As we intend to build a framework capable of describing network connectivity in full, i.e. network path diversity, we begin with a discussion of fundamental topological units that intrinsically embody a notion of the latter.

The simplest and smallest diversity topological unit is a {\em cycle} of nodes, as it affords two disjoint path choices to go from any node in the cycle to any other node in the same cycle. All paths in a network either lie on a part of the network which is a tree (no diversity) or can be expressed as an arc on some cycle (the most general end-to-end path can be a concatenation of the above two types of path, but this does not alter the ensuing discussion).

\begin{figure}[!t]
\centering
\includegraphics{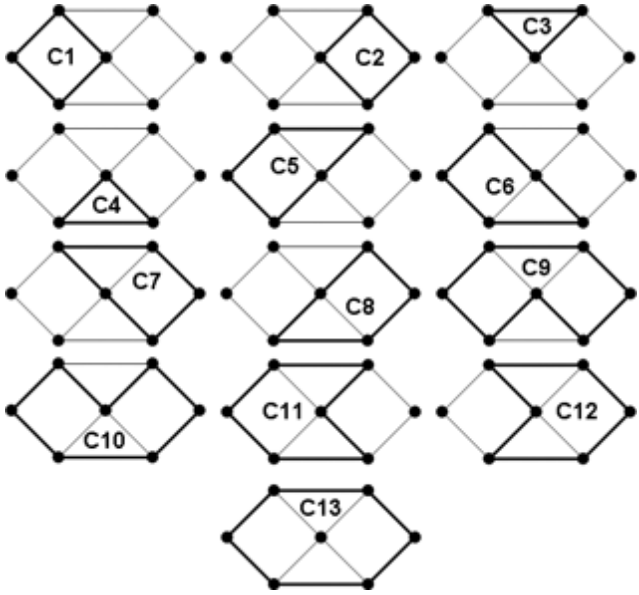}
\caption{The simple cycles of an example network, exhaustively enumerated.}
\label{Fig1}
\end{figure}

\begin{figure}[!t]
\centering
\includegraphics{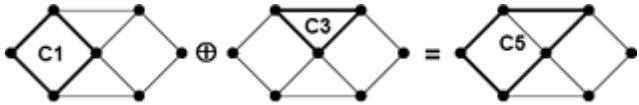}
\caption{Cycle algebra.}
\label{Fig2}
\end{figure}

As can be seen in the example of Figure~\ref{Fig1}, even a small, simple network graph can be decomposed into a fair number of cycles. However, not all $13$ cycles in this example are necessary to describe the graph fully. The size of the cycle space of a network of $n$ vertices and $m$ edges can become very large indeed, but this is not significant: A small subset of {\it independent} cycles can suffice in describing the cycle structure of a network. The notion of independence is intrinsically related to that of a linear vector space of cycles and necessitates the definition of an operator to combine cycles: The required operator is that of the symmetric difference between two cycles (this is the Boolean XOR operation on their edges), as can be seen in Figure~\ref{Fig2}. The maximal set of linearly independent cycles required to generate all the remaining cycles forms a basis set and the size, $\nu$, of the cycle basis set of a network is given by the elementary graph-theoretic result, $\nu = m - n + c$ where $c$ is the number of connected network components. The cycle basis set of a network is not unique and understanding the classes of cycle basis sets is currently the subject of intense research interest in the discrete mathematics community \cite{Liebchen2007}.

For example, the minimal cycle basis selects cycles that minimise the overall weight of the edges comprising the basis cycles. The computational complexity of finding the minimal cycle basis of a graph of $n$ vertices and $m$ edges is polynomial ($O(n m^3)$) for the simplest algorithm known \cite{Horton1987}, but faster versions exist (e.g. $O(m^3 + m n^2 \log n)$ \cite{Michail2006}).

The simple cycles in the basis set for a network can then be considered to be one of the fundamental units of path diversity in a network. The symmetric difference operator that enables us to combine cycles in the basis set in order to generate any other cycle yields a clue as to what is the second fundamental unit of path diversity: The Boolean AND operator between the edges of any two cycles in the basis set is either the null set of edges, or at least one edge (with its incident pair of vertices) and can be also seen to be a unit of {\it diverse connection between cycles}, distinct from the cycles themselves (note that a single vertex as an intersection between cycles is not a unit of path {\em diversity}). We shall henceforth refer to cycles as units of path diversity of type 1 and the diverse connections between cycles as units of path diversity of type 2. These two units are to replace the vertices and edges as the elementary objects comprising a network, if we are to build graph-theoretical constructs that are capable of supporting adaptive routing. More importantly, units of path diversity of type 1 can be thought of as blocks (Kadanoff transformation), whereas units of path diversity of type 2 can be thought of as interactions between the former.

\section{Logical Network Abridgment Procedure}

\begin{figure}[!t]
\centering
\includegraphics[width=1.8in]{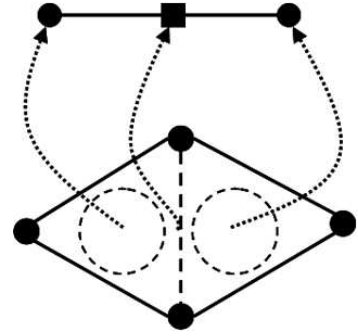}
\caption{Abstraction of cycles and cycle adjacencies into the two types of elementary units of diversity.}
\label{Fig3}
\end{figure}

\begin{figure}[!t]
\centering
\includegraphics{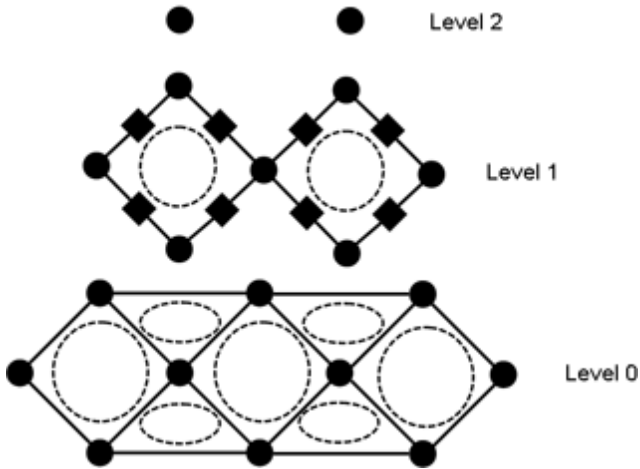}
\caption{The LNA of a simple network with a disjoint higher level.}
\label{Fig4}
\end{figure}

We can now construct a hierarchy of network abstraction graphs starting with the actual/physical network under consideration, as follows: At any level in the abstraction determine a cycle basis, according to some additional criteria suited to the application. Abstract all cycles in the basis as vertices of type 1 and all diverse cycle adjacencies as vertices of type 2 at the next level. Join vertices of type 1 to those vertices of type 2 which share at least 1 edge to form a bi-partite graph at the next level of abstraction, as can be seen in Figure~\ref{Fig3}. The process then repeats $L$ times until a cycle-free graph is arrived at, as in the example of Figure~\ref{Fig4}. We call the hierarchical ensemble of graphs the {\it logical network abridgment} (LNA) of the network \cite{LNA}. We enumerate the levels of abstraction $0, 1, \ldots, L$ for ease of reference.

Some remarks need to be made concerning graphs which contain trees as sub-graphs, cut vertices and cut edges. When a tree is attached to a cycle, we can collapse it {\it logically} into its vertex that is rooted in some cycle. Thus, trees do not appear in the graph of the next higher level. For the purposes of routing this simply signifies the fact that the vertex of the tree rooted in the cycle has the unique position of being responsible for forwarding data to all the destination vertices in the tree.

A network graph with a cut vertex or a cut edge becomes disjoint at the next level of LNA abstraction. This denotes the absence of diverse connectivity between vertices on either side of the cut vertex/edge at the lower level. Therefore, routing can only take place at the lower level of the LNA abstraction and the routing decision in this instance is unique, as it involves a deterministic forwarding decision at the cut vertex or vertices incident on the cut edge. It is of interest to point out that the cycle bases on either side of a cut vertex/edge decouple into two independent sub-spaces.

Both the number of levels of abstraction $L$, as well as the structure of the graph at each level of abstraction, contain information on how diversely connected a network is. Every level of abstraction conveys summarised path diversity information for the lower level, which can aid both the visualisation and analysis of this diversity. The summarisation is not done on an arbitrary clustering basis, but is dictated by the underlying network topology and introduces a natural measure for the network diversity, $L$. Clearly, the bigger $L$, the more intrinsic path diversity exists in a network. If the graph at any level of abstraction becomes disconnected, this indicates the existence of a path diversity bottleneck at the lower levels. An example of
the application of the LNA procedure to a graph illustrating the
above point is shown in Figure~\ref{Fig4}.

The convergence properties of the LNA are fairly well-understood, but will not be discussed here. At one extreme, for a connected network whose graph is a tree ($m=n-1$, $c=1$), $L=0$. At the other extreme, for a completely connected network of $n$ nodes (complete graph $K_n$) having $m=n(n-1)/2$ edges ($c=1$), it can be shown that $L=n-2$. Defining the path diversity density of a network to be $D \equiv L/n$, we can see that $0 \leq D \leq 1-2/n < 1$.

The path diversity density of a network can be used to determine the appropriateness of the choice of a routing technology to the specific topology of the network in question. If $D \simeq 0$, the network topology is dominated by trees and a protocol that computes shortest paths to each destination will be both highly scalable and efficient. In contrast, if $D \lesssim 1$, the network topology is close to being completely connected  and no ``intelligence"  is necessary in the routing technology: With a high degree of probability the destination vertex will be directly connected to the source vertex; if not, choosing a random adjacent vertex also has a high probability of this being directly connected to the destination vertex. As a consequence, random deflection routing will deliver the data to its destination with a probability that approaches unity exponentially fast in the number of hops required for successful delivery. In this instance random deflection routing is scalable, robust and sufficient.

The intermediate $D$ case ($0 < D \ll 1$) is more problematic. A shortest path routing protocol fails to exploit the underlying network diversity and will take time to re-converge if congestion or failures arise. Random deflection routing is unlikely to deliver data successfully to its destination, as nodes are likely to be separated by many hops (in fact the probability of successful delivery becomes exponentially small in the number of hops along the shortest path tree that separates the source and destination vertices). Therefore, in order to exploit the underlying network diversity a dynamic, adaptive routing protocol is required. As we shall see shortly, the natural framework on which to base the creation of such a protocol is the LNA itself.

\section{Resilient Recursive Routing}

The LNA can be augmented with a number of forwarding rules to create a resilient recursive routing (R$^3$) protocol. A routing algorithm capable of operating efficiently in the intermediate $D$ case must exploit the LNA and operate recursively at each level of abstraction of the network, either to route a packet around a single cycle, or along a tree. Routing information on a tree is a trivial exercise, in the sense that all forwarding decisions are deterministic and we shall not discuss this any further. The fundamental generic algorithm must route a packet from a source to a destination, both of which are members of the same level $1$ logical node and thus are members of the same cycle at level $0$ (hereafter referred to as level $0$ neighbours). The algorithm must be capable of (i) loop-free data routing across the cycle, (ii) load balancing across the cycle and (iii) fast reaction to link or node failures in the level $0$ cycle of nodes.

If the source and destination are members of the same level $2$ logical node (i.e. they belong to the same level $1$ cycle and are thus level $1$ neighbours), the fundamental routing algorithm should be applied iteratively twice, once at level $1$ and once at the current (local) level $0$ cycle.

\begin{figure}[!t]
\centering
\includegraphics{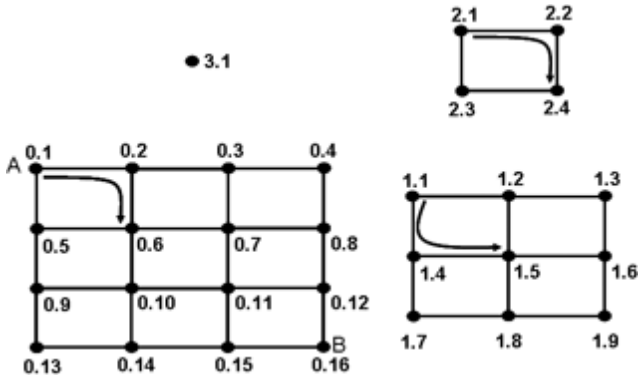}
\caption{Resilient Recursive Routing.}
\label{Fig5}
\end{figure}

For source and destination nodes that are level $\ell$ neighbours, the fundamental routing algorithm needs to be applied $\ell+1$ times iteratively, from the current highest level $\ell$ down to the local level $0$ cycle, as illustrated by Figure~\ref{Fig5}.

If at some level of abstraction $\ell'$ the LNA graph of the network is disjoint (in Figure~\ref{Fig4} for example $\ell'=2$), the fundamental routing algorithm cannot find a level $1$ cycle or tree across some source and destination pairs. In this case, the algorithm must drop down to level $\ell'-1$, where at least one cut-node (as shown in Figure~4) needs to be traversed {\em deterministically} at the $\ell'-1$ level of abstraction, just as routing on a tree needs to operate. This implies that cut-nodes need to exchange reachability information about their corresponding bi-connected parts of the network.

The routing methodology embodied in the generic algorithm must enable us to route a packet in a loop-free manner, while performing load balancing and enabling failure recovery across the network. The iterative nature of the algorithm though does not on its own guarantee the scalability of all the properties of the fundamental routing algorithm to the entire network. The first condition necessary for the scalability of the routing protocol is the need to have the number of levels of abstraction $L$ to be significantly smaller than the number of nodes $n$ in the original network, as the size of the network grows, i.e. $L \ll n$, or equivalently $D \ll 1$. The second condition necessary to achieve protocol scalability relates to the characteristic reaction times of the fundamental routing algorithm to congestion and failures at the higher levels of abstraction. The higher levels must use summarised information, e.g. for congestion along their logical cycles, over longer time-scales to reflect the summarised nature of this higher-level topological neighbourhood. For example, if for a sufficiently sparse class of network graphs it were to turn out that $L \sim \log n$, as $n \to \infty$, it would be natural to select adaptation/update time intervals, $\tau_{\ell}$, for higher levels that grow exponentially, $\tau_{\ell} \approx \tau_0 \cdot b^\ell, \ \ell = 0, \ldots, L$, for some base $b > 1$ which depends on the sparsity of the graph and a desirable fastest adaptation time, $\tau_0$, at physical level $\ell=0$.

\section{Conclusion}

The routing protocol we present can itself be considered to be analogous to the interaction mechanism of source and destination node pairs. Scalability of the protocol properties is achieved through its recursive nature, which is essentially the algorithmic expression of the Kadanoff transformation.

The generic routing algorithm introduces natural metrics that lead to the definition of topological distance (or neighbourhood), namely the number of the logical level at which two physical nodes belong to the same logical vertex (or equivalently the logical cycle one level below this). The subsequent association of a range of topological distances in a network with a corresponding range of time-scales can then become the building block for a truly adaptive family of routing protocols. The maximum topological distance, $L$, leading to the normalised topological diameter, $D=L/n$, is also an important parameter that determines the class of networks for which LNA-based routing protocols can be made scalable.

Naturally, numerous open questions remain. The most significant example of such an open question is how can stable routing be achieved with incomplete or partially inconsistent topological information. Numerous graph theoretic issues also arise in the context of making the LNA unique and its computation efficient.

Besides its application to routing, the LNA lends itself to the analysis of the structure of complex networks in terms of vulnerability to attrition of nodes, links and common failure groups of nodes and links.

\section*{Acknowledgment}

The authors would like to thank the UK Engineering and Physical Sciences Research Council for supporting this work under grant GR/T23725/01.

\end{document}